\documentclass[11pt]{article}

\usepackage{amssymb}
\usepackage{amsthm}
\usepackage{amsmath}
\usepackage[noend]{algorithmic}
\usepackage{algorithm}
\usepackage{pgf,tikz}
\usetikzlibrary{arrows}

\def\pspace{\mbox{\textsc{PSpace}}}
\def\pclass{\mbox{\textsc{P}}}
\def\dynamics{\mathbf{d}}
\def\arcdynamics{\mathbf{a}}

\newtheorem{lemma}{Lemma}
\newtheorem{theorem}{Theorem}

\newcommand{\virgole}[1]{``#1''}

\begin{document}








\begin{center}
{\Large{\bf Opinion Evolution among friends and foes: the deterministic Majority Rule
- extended abstract}}

\bigskip

{\large Miriam Di Ianni}

\medskip

{\small \em Dipartimento di Ingegneria dell'Impresa \virgole{Mario Lucertini} \\Universit\`a degli Studi di Roma \virgole{Tor Vergata}\\Roma, Italy. E-mail: miriam.di.ianni@uniroma2.it}
\end{center}

\begin{abstract}
The influence of the social relationships of an individual on the individual's opinions (about a topic, a product, or whatever else) is a well known phenomenon and it has been widely studied. This paper considers a network of positive (i.e. trusting) or negative (distrusting) social relationships 
where every individual
has an initial positive or negative opinion (about a topic, a product, or whatever else) that changes over time, at discrete time-steps, due to the influences each individual gets from its neighbors. 
Here, the influence of a trusted neighbor is consistent with the neighbor's opinion, while the influence of an untrusted neighbor is opposite to the neighbor's opinion. This extended abstract introduces the {\em local threshold-based} opinion dynamics and, after stating the computational complexity of some natural reachability problems arising in this setting when individuals change their opinions according to the opinions of the majority of their neighbors, proves an upper bound on the number of opinion configurations met by a symmetric positive-only relationships network evolving according to any of such models, which is polynomial in the size of the network. This generalizes a result in \cite{Chatterjee2020}.
\end{abstract}

\section{Introduction}
The fact that the network of social relations of an individual influences the individual's behavior - opinions, purchases, voting - is a well known phenomenon and has been studied in several contexts. In particular, the Influence Maximization, aiming at detecting a fixed set size of individuals able to maximize the spread of an opinion, and the Target Set Selection, aiming at detecting the minimum size set of individuals able to convince all the other individuals in the network,
have received a wide attention in the literature and some mathematical models have been proposed to analyze such a kind of diffusion of information over networks, including the linear threshold model \cite{Granovetter1978}, the voter model \cite{Holley1975} and the Independent Cascade Model \cite{GLM2001,KKT2003}, all of which considering relationships inducing {\em positive} feedback effects only.

However, in most network settings also {\em negative} link effects are to be considered. 
One way to take into account negative feedbacks is that pursued in
\cite{Chen2011, NazTag2012, Stitch2014, YaWaTru2019} 
where it is assumed that individuals may also develop a negative opinion about the feature to be spread and, in this case, they may negatively influence their neighbors. 
Differently, in
\cite{Ahmed2013, DV2020, LiChen2013} the possibility that some relations between individuals are ruled  by, for instance, antagonism and distrusting (see \cite{EaKlein2010} and references quoted therein) is considered: it is now assumed that node opinions about the feature to be spread are always positive but that receiving positive feedback from an untrusted/antagonist neighbor results in increasing the support to discard the feature. Finally, in
\cite{ GaArRoy2016, HZND2016} the two approaches (positive/negative opinions and positive/ne\-ga\-tive relationships) are jointly considered.

Strictly related to the analysis of the diffusion of information is the analysis of opinion dynamics, in which  individuals have a state (that is, an opinion) which evolves over time~\cite{Sirbu2017}. Now, the questions under considerations are something like the following: starting at some given current opinion configuration, will an equilibrium opinion configuration or a given target opinion configuration  ever be reached? Or, also, will a given target individuals set ever reach consensus? Different opinion dynamics stochastic models have been considered as to unsigned relations (such as, the French-DeGroot model~\cite{French1956,Degroot1974} and its extensions, the majority rule model~\cite{Galam2002}, and the social impact model~\cite{Lewenstein1992}) and some of these models have been adapted to the case of signed graphs~\cite{Li2015,Shi2016,He2021,Xue2020}. In \cite{Chatterjee2020} a couple of deterministic opinion dynamics models are defined, as a simplification of the Game-of-Life~\cite{Gardener1970}, and their reachability properties are studied. One of the two models, the {\em underpopulation} opinion dynamics, will be furtherly considered in this paper.

\subsection*{Paper contributions}
Letting individuals change their opinion on a majority basis with respect to their neighbors' opinions is a quite natural assumption which has indeed somehow taken into accounts in some of the proposed models. As an example, in the Voter model \cite{CS1973,Moretti2013} each individual has one of two discrete opinions and at each time step a random individual is selected along with one of its neighbours with the first one taking the opinion of the neighbour: here, majority plays an indirect role in that an individual has a greater probability to get the opinion of the majority of its neighbors than the opposite one. The role of the majority's opinion is made explicit in the Majority Rule model first proposed in \cite{Galam2002}. In \cite{Galam2002} agents take discrete opinions ($+1$ or $-1$) and can interact with all other agents (that is, an underlying complete unsigned graph is considered); at each time step a group of $r$ agents is randomly selected and all of them take the majority opinion within the group.

In this paper the {\em deterministic Majority Rule} opinion dynamics is introduced and studied, which can be considered as the deterministic counterpart of the Majority Rule model proposed in \cite{Galam2002}. 
In the Deterministic Majority Rule individuals operate in an underlying directed signed (non-complete) graph, where an incoming positive (negative) arc to an individual describes that the individual trusts (respectively, distrusts) that neighbor; at each time step each individual takes the majority opinion within the set of its neighbors, where, like in \cite{Li2015}, an opinion passing through a negative arc is complemented (that is, a distrusted neighbor with a negative opinion is equivalent to a trusted neighbor with a positive opinion, and a distrusted neighbor with a positive opinion is equivalent to a trusted neighbor with a negative opinion). 

Three reachability-related problems are here considered with respect to the Deterministic Majority rule: given a signed graph in a given {\em opinion configuration} (that is, an assignment of a positive or negative opinion to each of its nodes), will an equilibrium opinion configuration ({\sc ReachEquilibrium}) / a given target opinion configuration ({\sc Reachability}) / an opinion configuration in which all nodes in a given target set agree ({\sc ReachTarget}) ever be reached?

The achievement of this paper is showing that, while the link signs are ininfluent to the complexity of the aforementioned reachability problems, such complexity dramatically changes when considering directed or undirected graphs. This is formalized in the following theorem.


\begin{theorem} \label{thm::reachdirected}
The problems {\sc Reachability}, {\sc ReachTarget} and {\sc ReachE\-qui\-librium} considered with respect to the Deterministic Majority Rule   are \pspace-complete even when restricted to unsigned directed graphs.
\end{theorem}
\begin{theorem} \label{thm::reachundirected}
The problems {\sc Reachability}, {\sc ReachTarget} and {\sc ReachE\-qui\-librium} are in \pclass\ when restricted to signed undirected graphs.
\end{theorem}

The proof of the above theorems is deferred to the full version of this paper. Instead, in this extended abstract the generalization of a result  in \cite{Chatterjee2020} which is functional to the proof of Theorem \ref{thm::reachundirected} will be formally stated and proved.

The proof of Theorem \ref{thm::reachundirected} strongly relies on the fact that the number of opinion configurations met by an undirected unsigned graph during its opinion evolution occurring with respect to the Majority Deterministic rule and starting at any initial opinion configuration is polynomially bounded on the size of the graph. A similar polynomial bound had already been proved in \cite{Chatterjee2020} in the case in which an undirected unsigned graph evolves according to the {\em underpopulation} rule defined in that paper as a simplification of the Game-of-Life rule. 
Actually, both the deterministic Majority Rule and the underpopulation rule fall within a general framework rule introduced in this paper that will be referred to as {\em local threshold-based opinion dynamics} rule: for a given pair of computable integer threshold functions $\theta^+$ and $\theta^-$, at any step each individual decides whether changing its opinion or not based on the values $\theta^+(k)$ and $\theta^-(k)$, $k$ being the number of the individual's neighbors. As it will be described in Section \ref{sec::preliminaries}, the deterministic Majority Rule occurs when $\theta^+(k)$ and $\theta^-(k)$ are close to $k/2$, and  the underpopulation rule occurs when $\theta^+(k)=i_1$ and $\theta^-(k)=i_2$, for some pair of constants $i_1$ and $i_2$. 

Specifically, after having provided the needed definitions in Section  \ref{sec::preliminaries}, in Section \ref{sec::mainlemmaproof} the following theorem will be formally stated and proved.

\medskip

\noindent
{\bf Theorem 3.} {\em For any undirected unsigned graph $G=(V,E)$ of maximum degree $\Delta$ and for any initial opinion configuration $\omega$ of $G$, the number of opinion configurations met by $G$ during its opinion evolution starting at $\omega$ and occurring with respect to any local threshold-based dynamics rule is at most}
$$ 4|E|+2|V|+4 \Delta (\Delta +1)|V|+2.$$

\section{Preliminary definitions and notations} \label{sec::preliminaries} 

A directed {\em signed} graph $G=(V,A, \lambda)$ is a directed  graph together with an arc-labeling function $\lambda: A \rightarrow \{-1,1\}$. An undirected signed graph $G=(V,E, \lambda)$, is similarly defined with $\lambda$ being an edge labeling function, that is,  $\lambda: E \rightarrow \{-1,1\}$. 

Within this paper, for any node $v$ of a directed (undirected) signed graph $G$, $N(u)$ denotes the set of of in-neighbors (respectively, neighbors) of $u$.

An \textit{opinion configuration} of a signed graph $G$ is a node-labeling function $\omega: V \rightarrow \{-1,1\}$, stating whether a given node is in favor or against a specific topic. Nodes influence each other so that their opinions change over time. In particular, the neighbors of a node $u$ influence the opinion $u$ gets over time: positive in-neighbors positively influence $u$, that is, their influence works in favour of $u$ getting their same opinion, while negative in-neighbors negatively influence $u$, that is, their influence works in favour of $u$ getting their opposite opinion. In this respect, for any in-neighbor $v$ o any nodef $u$, we say that $v$ {\em pushes} $u$ to 1 at $\omega$  if $\omega(v)=1$ and $\lambda(v,u)=1$ or $\omega(v)=-1$ and $\lambda(v,u)=-1$, and that
$v$ {\em pushes} $u$ to $-1$ at $\omega$ if $\omega(v)=-1$ and $\lambda(v,u)=1$ or $\omega(v)=1$ and $\lambda(v,u)=-1$.

An \textit{opinion dynamics} is a functional $\dynamics$ which specifies, for a given signed graph $G$ and an opinion configuration $\omega$ of $G$, the next opinion configuration $\dynamics(G,\omega)$ of $G$. 
The \textit{opinion configuration evolution set} of a signed graph $G$ in a configuration $\omega$ with respect to an opinion dynamics $\dynamics$ (or, in short, the $\dynamics$-{\em evolution set} of $G$) is the sequence $\mathcal{E}_{\dynamics}(G,\omega)= \langle \omega_1=\omega, \omega_2, \ldots , \omega_T \rangle$ of distinct opinion configurations such that $T \leq 2^{|V|}$ (the number of configurations of $V$) and
\begin{itemize}
\item for $t=2, \ldots ,T$, $\omega_t=\dynamics(G,\omega_{t-1})$, and
\item there exists $h \leq T$ such that $\omega_h=\dynamics(G,\omega_{T})$.
\end{itemize}
Whenever it happens that $h=T$, $\omega_T$ is said an {\em equilibrium configuration}. 

With a slight abuse of notation, the sequence $\mathcal{E}_{\dynamics}(G,\omega)$ shall be dealt with as a set as well.

This paper focuses on the \textit{deterministic majority} opinion dynamics $\dynamics_M$
in which a node with positive (respectively, negative) opinion changes its opinion only if more than half of its in-neighbors push it to $-1$ (respectively, 1). 
Formally, for any node $u$ and for any opinion configuration $\omega$, $\dynamics_M(G,\omega)=\omega'$ is defined as 
\[\omega'(u) = \left\lbrace
\begin{array}{ll}
1 & \mbox{if $\sum_{v \in N(u)}\lambda(v,u)\omega(v)>0$}, \\
\omega(u) & \mbox{if $\sum_{v \in N(u)}\lambda(v,u)\omega(v) = 0$}, \\
-1 & \mbox{if $\sum_{v \in N(u)}\lambda(v,u)\omega(v) < 0$}.
\end{array}
\right.
\]

A dynamics $\dynamics$ is \textit{local} when the new opinion of a node depends only on its current opinion, the current opinions of its in-neighbors, and the signs of the arcs connecting them to it. A dynamics $\dynamics$ is \textit{threshold-based} when it is ruled by a pair of {\em threshold functions} $\theta^+: \mathbf{N} \rightarrow \mathbf{N}$ and $\theta^-: \mathbf{N} \rightarrow \mathbf{N}$ in the following way: for any node $u$ and for any opinion configuration $\omega$,  $\dynamics(G,\omega)=\omega'$ is defined as
\[\omega'(u) = \left\lbrace
\begin{array}{ll}
 1 & \mbox{if $\omega(u)=1$ and $P(u)\geq \theta^+(|N(u|))$} \\
  & \mbox{or $\omega(u) =-1$ and $P(u)\geq \theta^-(|N(u)|)$}, \\
-1 & \mbox{if $\omega(u)=1$ and $P(u)< \theta^+(|N(u)|)$} \\
 & \mbox{or $\omega(u) =-1$ and $P(u)< \theta^-(|N(u)|)$,}
\end{array}
\right.
\]
where $P(u)$ is the number of in-neighbours $v$ of $u$ pushing $u$ to 1. Notice that, without loss of generality, the bounds $\theta^+(u) \leq \Delta+1$ and $\theta^-(u) \leq \Delta+1$ for any $u \in V$ can be assumed, where $\Delta$ is the maximum node in-degree in $G$.

It can be easily verified that the deterministic Majority Rule is the local threshold-based dynamics corresponding to having $\theta^+(k)=\left\lceil\frac{k}{2}\right\rceil$ and  $\theta^-(k)=\left\lfloor\frac{k}{2}\right\rfloor+1$.
The \textit{underpopulation} opinion dynamics considered in \cite{Chatterjee2020} is another noticeable example of a  local threshold-based opinion dynamics which corresponds to having $\theta^+(k)=t_{1}$ and  $\theta^-(k)=t_{2}$ for some pair of constants $t_{1},t_{2}\in\mathbf{N}$.

\section{Local threshold-based dynamics in undirected unsigned graphs: size of the opinion configuration evolution set} \label{sec::mainlemmaproof}

Aim of this subsection is proving that, for any local threshold-based dynamics $\dynamics$, the size of $\mathcal{E}_{\dynamics}(G,\omega)$ is polynomially bounded by the size of $G$. This will be accomplished by exploiting
the same reasoning applied in \cite{Chatterjee2020} for proving the same result in the case of the underpopulation rule, properly modified to let it works for a generic local threshold-based dynamics.
%

For the sake of readability (and for keeping the notation as close as possible to that in \cite{Chatterjee2020}), within this section the opinion $-1$ will be replaced by $0$, that is, we shall assume that an opinion configuration is a function $\omega: V \rightarrow \{ 0,1 \}$; needless to say, this does not change anyway the model.
%

Let $G=(V,E)$ be an undirected unsigned graph, let $\omega$ be an opinion configuration of $G$ and let $\mathcal{E}_{\dynamics}(G,\omega)= \langle \omega_1=\omega, \ldots , \omega_T \rangle$ be the $\dynamics$-evolution set of $G$, where $\dynamics$ is any local threshold-based dynamics. For $v \in V$, the {\em history} 
of $v$ is the 
string $\omega_1(v) \omega_2(v) \ldots \omega_T(v) \in \{ 0,1 \}^T$, that is, the history of $v$ is the sequence of opinions that $v$ gets during the $\dynamics$-evolution of $G$ starting at $\omega$ before an opinion configuration is repeated (recall that $\dynamics(\omega_T)=\omega_h$ for some $h \leq T$). For $1 \leq i \leq j \leq T$, the {\em period} $\omega_{[i,j]}(v)$ in the history of $v$ is the string $\omega_i(v) \omega_{i+1}(v) \ldots \omega_j(v)$.

In what follows, a sequence $y= y_1y_2 \ldots y_h \in \{ 0,1,? \}^h$ will be used as a shortcut to the set of all sequences in which every ? inside $y$ is replaced by 0 and by 1: as an example, $?1?$ stands for the set of sequences $010, 011, 110, 111$. Given a pair of sequences $y= y_1y_2 \ldots y_h \in \{ 0,1,? \}^h$ and $z=z_1z_2 \ldots z_h \in \{ 0,1 \}^h$,  $y$ {\em matches} $z$ (in symbols, $y \approx z$) if $y_i=z_i$ whenever $y_i \in \{ 0,1 \}$ (no matter what happens if $y_i=?$), for every $i=1, \ldots ,h$. 

Given a node $v$ and a sequence $y\in\{0,1,?\}^h$, let $[y,v]$ be the number of matches of $y$ inside the history of node $v$, that is, 
\[
[y,v] = \big\lvert \{ i \in \{ 1, \ldots , T-h+1 \}: y \approx \omega_{[i,i+h-1]}(v) \} \big\rvert ,
\]
and let $[y] = \sum_{v \in V} [y,v]$ be the total number of matches of $y$ inside the histories of all nodes in $V$.
Hence, next lemma shows that the total number of matches of 110, 100, 011 and 001 inside the histories of all nodes in $V$ is upper-bounded by a polynomial in the size of $G$.
\begin{lemma} \label{lemma::1?0+0?1}
For any undirected unsigned graph $G$ and for every $t \geq 3$, 
\[ [110]+[100]+[011]+[001] \leq 4|E|+2|V| + 4|V| \Delta (\Delta +1), \]
where $\Delta$ is the maximum node degree in $G$.
\end{lemma}
\begin{proof}
For any $y= y_1y_2 \ldots y_h \in \{ 0,1,? \}^h$, with $h \leq T$, denote as $[y]^1$  the number of nodes whose history starts with the sequence $y$ and as $[y]^T$ the number of nodes whose history ends with the sequence $y$. Hence, $0 \leq [y]^1 \leq |V|$, $0 \leq [y]^T \leq |V|$.

The first step is bounding $[001] + [011]$. To this aim, notice that, trivially, $[00] = [00]^1 + [000] + [100]$ and $[00]=[00]^T + [000] + [001]$; from the two equalities it follows that
\begin{equation} \label{eq::from00_t}
| [100] - [001] | = | [00]^1 - [00]^T | \leq |V| ;
\end{equation} 
similarly, $[11] = [11]^1 + [011] + [111]$ and $[11]=[11]^T + [110] + [111]$, so that
%
\begin{equation} \label{eq::from11_t}
| [011] - [110] |  \leq |V|.
\end{equation} 
Hence,
 \begin{equation} \label{eq::0?1}
 [001] + [011] 
   \ \leq \ [100] + |V| + [110] +|V| \ = \ [1?0] + 2 |V|.
\end{equation}

In order to bound $[1?0]$ some more notation is needed. First of all, for any $1 \leq k \leq \Delta$,  $V_k$ is the subset of $V$ containing all the nodes of degree $k$. 

For $y \in \{ 0,1,? \}^*$ such that $|y| \leq T$, $[y,k]$ denotes the total number of matches of $y$ inside the history of all nodes in $V_k$ with $[y] = \sum_{k=1}^{\Delta} [y,k]$, since $\{ V_k \}_{1 \leq k \leq \Delta}$ is a partition on $V$; $[y,k]_t^T$ and $[y,k]^0$ are defined accordingly.

Let $y,z \in \{ 0,1,? \}^*$ be such that $|y|=|z| \leq T$. Then, 
\begin{small}
\[ [y,z,k] =  \sum_{u \in V_k} \sum_{v \in N(u)} | \{ i \in \{ 0, \ldots, T-|y| \} :
\omega_{[i+1,i+|y|]}(u) \approx y \ \wedge \ \omega_{[i+1,i+|z|]}(v) \approx z \}| \]
\end{small}
is the number of corresponding matches of $y$ and $z$ inside the histories  of any pair of nodes $u$ and $v$ such that $(u,v) \in E$ and $u \in V_k$. Similarly, 
\begin{small}
\[ [y,z] = \sum_{u \in V} \sum_{v \in N(u)} | \{ i \in \{ 0 \ldots T-|y| \}:  \omega_{[i+1,i+|y|]}(u) \approx y \ \wedge \ \omega_{[i+1,i+|z|]}(v) \approx z \}| \]
\end{small}
is the number of corresponding matches of $y$ and $z$ inside the histories of all  pair of nodes $u$ and $v$ such that $(u,v) \in E$. Since $\{ V_k \}_{1 \leq k \leq \Delta}$ is a partition on $V$, then $[y,z] = \sum_{k=1}^{\Delta} [y,z,k]$. 

Finally, 
\[ [y,z,k]^0 = \sum_{u \in V_k \ : \ \omega_{[1,|y|]}(u) \approx y} \left|  \{ v \in N(u):  \ \omega_{[1,|z|]}(v) \approx z \}  \right| \]

and $[y,z]^0$, $[y,z,k]^T$ and $[y,z]^T$ are defined similarly.

Notice that, by the edge simmetry, for every $y,z \in \{ 0,1,? \}^*$ such that $|y|=|z| \leq T$, it holds that $[y,z] = [z,y]$ and so it is $[?1, 1?] = [1?, ?1]$. Let us now compute in two different ways the two sides of the last equality:
\begin{eqnarray*} 
[?1, 1?] & = & [?1, 1?]^1 + [0?1, ?1?] + [1?1, ?1?] \\
 & = & [?1, 1?]^1 + [001, ?1?] + [011, ?1?] + [1?1, ?1?] 
 \end{eqnarray*} 
 \begin{eqnarray*} 
[1?, ?1] & = & [1?, ?1]^T + [1?0, ?1?] + [1?1, ?1?] \\
 & = & [1?, ?1]^T + [100, ?1?] + [110, ?1?] + [1?1, ?1?].
\end{eqnarray*} 
Hence, 
by equalizing the last terms in the two chains of equalities above, 
$$ | [001, ?1?] + [011, ?1?] - [100, ?1?] - [110, ?1?] | = | [?1, 1?]^1 - [1?, ?1]^T |. $$
Similarly as before, it holds that $0 \leq [?1, 1?]^1 \leq 2|E|$ and $0 \leq [?1, 1?]_t^T \leq 2|E|$, so that
\begin{equation} \label{eq::edges1}
 -2|E| \leq [001, ?1?] + [011, ?1?] - [100, ?1?] - [110, ?1?] \leq 2|E|.
\end{equation}
As far as $[001, ?1?] , [011, ?1?] , [100, ?1?] $ and $[110, ?1?]$ are concerned, by the definition of $\dynamics$, the following holds. Since the state of a node $u$ changes from $0$  to $1$ if and only if at least $\theta^-(|N(u)|)$ of its neighbors are in state $1$ and since a node $u$ in state $0$  remains in  state $0$ if and only if less than $\theta^-(|N(u)|)$ of its neighbors are in state $1$, then
$$[001, ?1?,k] \geq \theta^-(k) [001, k]_t \mbox{\hspace*{0.1cm} and \hspace*{0.1cm}}[100, ?1?,k]  \leq \left( \theta^-(k)-1\right)  [100, k]_t $$
and, similarly,  
$$[011, ?1?,k] \geq \theta^+(k) [011, k]_t \mbox{\hspace{0.1cm} and \hspace*{0.1cm}} 
[110, ?1?,k] \leq \left( \theta^+(k) -1 \right) [110, k]_t. $$

Hence, 
\begin{small}
\begin{align*}
 & [001, ?1?] - [100, ?1?] + [011, ?1?] - [110, ?1?]  = \\
& \sum_{k =1}^{\Delta} \left\lbrace [001, ?1?,k] - [100, ?1?,k] + [011, ?1?,k] -  [110, ?1?,k] \right\rbrace \geq\\
& \sum_{k =1}^{\Delta} \left\lbrace \theta^-(k) [001,k]  -  \left( \theta^-(k)-1\right)  [100,k] 
		 	+ \theta^+(k)  [011,k] - \left( \theta^+(k) -1 \right) [110,k] \right\rbrace =  \\
&  \sum_{k =1}^{\Delta} \left\lbrace \theta^-(k) ([001,k]  -  [100,k]) +   [100,k] 
			+ \theta^+(k)  ([011,k] - [110,k]) + [110,k] \right\rbrace =  \\
&   [100]  +  [110] + \sum_{k =1}^{\Delta} \left\lbrace \theta^-(k)( [001,k]  -   [100,k]  )  
			+ \theta^+(k)  ([011,k] -  [110,k]) \right\rbrace.\\
\end{align*}
\end{small}
And, by (\ref{eq::edges1}), this implies 
\begin{small}
\begin{align*}
    [100]  +  [110] &\leq 2|E| +\sum_{k =1}^{\Delta} \left\lbrace  \theta^-(k) ( [100,k]  -   [001,k]  ) 
    			+ \theta^+(k)   ([110,k] -  [011,k] \right\rbrace. \\
\end{align*}
\end{small}
By the same reasonings to those leading to Equations (\ref{eq::from00_t}) and (\ref{eq::from11_t}), it holds that $|[100,k] - [001,k]| \leq |V|$ and $|[110,k] - [011,k]| \leq |V|$. Hence, by recalling that $\theta^+(u) \leq \Delta+1$ and $\theta^-(u) \leq \Delta+1$,
%
\begin{small}
\[ [100]  +  [110] \leq 
   2|E| +|V| \sum_{k =1}^{\Delta} \left\lbrace \theta^-(k)+\theta^+(k) \right\rbrace
 \leq 2|E| + |V| \Delta (2\Delta +2).
\]
\end{small}

Finally, by (\ref{eq::0?1}), the assertion follows
\end{proof}

The next theorem then follows from the above Lemma.
\begin{theorem} \label{thm::undirectedstable-lengthoftransient}
For any local threshold-based opinion dynamics $\dynamics$, for any undirected unsigned graph $G=(V,E)$ and for any opinion configuration $\omega$ of $G$,
$\left| \mathcal{E}_{\dynamics}(G,\omega) \right| \leq 4|E|+2|V|+4 \Delta (\Delta+1) |V| +2$.
\end{theorem}
\begin{proof}
Let $T=\left| \mathcal{E}_{\dynamics}(G,\omega) \right|$. Since the opinion configurations in $\mathcal{E}_{\dynamics}(G,\omega)$ are distinct, then, in particular,  for any $1 \leq t \leq T-2$, $\omega_t \neq \omega_{t+2}$. This means that, for any $1 \leq t \leq T-2$, there exists $u_t \in V$ such that $\omega_t(u_t) \neq \omega_{t+2}(u_t)$. 
As a consequence, for any $1 \leq t \leq T-2$, the string $\omega_t(u_t) \omega_{t+1}(u_t) \omega_{t+2}(u_t)$ is one in the set $\{ 001,011,100,110 \}$.

Hence, $[001]+[011]+[100]+[110] \geq T-2$ and, 
by Lemma \ref{lemma::1?0+0?1},
$$T-2 \leq 4|E|+2|V|+4 \Delta (\Delta+1) |V|. $$
\end{proof}

\bigskip

\noindent
{\bf Aknowledgements.} The author wishes to thank Pierluigi Crescenzi an Giorgio Gambosi for fruitful discussions and for having read preliminary versions of this paper.

\bibliographystyle{elsarticle-num}

\begin{thebibliography}{00}
\bibitem{Ahmed2013} S. Ahmed, and C.I. Ezeife.  Discovering influential nodes from trust network. In{\em  Proceedings of the 28th Annual ACM Symposium on Applied Computing}, 121--128, 2013.

\bibitem{Chatterjee2020} K. Chatterjee, R. Ibsen{-}Jensen, I. Jecker, and J. Svoboda. Simplified Game of Life: Algorithms and Complexity. In {\em Proceedings of 45th International Symposium on Mathematical Foundations of Computer Science}, 22:1--22:13, 2020.

\bibitem{Chen2011} W. Chen, A. Collins, R. Cummings, T. Ke, Z. Liu, D. Rincon, X. Sun, Y. Wang, W. Wei, and Y. Yuan. Influence maximization in social networks when negative opinions may emerge and propagate. In {\em Proceedings of the 2011 SIAM international conference on data mining},  379--390, 2011.

\bibitem{CS1973} P. Clifford, and A. Sudbury. A model for spatial conflict. {\em Biometrika} 60(3), 581--588, 1973.

\bibitem{Degroot1974} M.H. Degroot. Reaching a Consensus. {\em Journal of the American Statistical Association}, 69, 118--121, 1974.
\bibitem{DV2020} M. Di Ianni, and G. Varricchio.
Latency-Bounded Target Set Selection in Signed Networks. {\em Algorithms} 13(2): 32, 2020.


\bibitem{EaKlein2010} D. Easley, and J. Kleinberg.  Networks, Crowds, and Markets: Reasoning about a Highly Connected World. {\em Significance}, 9, 43--44, 2012.

\bibitem{French1956} J.R.P. French. A formal theory of social power. {\em Psychological review}, 63(3), 181--194, 1956.

\bibitem{Galam2002} S. Galam. Minority opinion spreading in random geometry. {\em The European Physical Journal B - Condensed Matter and Complex Systems}, 25, 403--406, 2002.
 
\bibitem{GaArRoy2016} S. Galhotra, A. Arora, and S. Roy.  Holistic influence maximization: Combining scalability and efficiency with opinion-aware models. In {\em Proceedings of the 2016 International Conference on Management of Data}, 743--758, 2016.

\bibitem{Gardener1970} M. Gardener. Mathematical games: the fantastic combinations of John Conway's new solitaire game ``life''. {\em Scientific American}, 223(4), 120--123, 1970.

\bibitem{GLM2001} J. Goldenberg,  B. Libai, and E. Muller.
Talk of the network: A complex systems look at the underlying process of word-of-mouth. {\em Market. Lett.}, 12(3), 211--223, 2001.

\bibitem{Granovetter1978} M. Granovetter. Threshold models of collective behavior. \textit{American Journal
of Sociology}, 83(6), 1420–1443, 1978.

\bibitem{He2021} G. He, H. Ruan, Y. Wu and J. Liu. Opinion Dynamics With Competitive Relationship and Switching Topologies. {\em IEEE Access}, 9, 3016--3025, 2021.

\bibitem{Holley1975} R. A. Holley,  and T. M. Liggett. Ergodic theorems for weakly interacting infinite systems and the voter model.\textit{The Annals of Probability}, 3(4), 643–663, 1975.

\bibitem{HZND2016} M. Hosseini-Pozveh, K. Zamanifar, A.R. Naghsh-Nilchi, and P. Dolog.  Maximizing the spread of positive influence in signed social networks. {\em  Intell. Data Anal.} 20(1), 199--218, 2016.

\bibitem{KKT2003} D. Kempe, J. Kleinberg, and \'{E}. Tardos.
Maximizing the Spread of Influence Through a Social Network.
In \textit{Proceedings of the Ninth ACM SIGKDD International Conference on Knowledge Discovery and Data Mining}, Washington, DC, USA, pp. 137--146, 2003.


\bibitem{Lewenstein1992} M. Lewenstein,  A. Nowak, and B. Latan\'e. Statistical mechanics of social impact. {\em Phys. Rev. A}, 45(2), 763--776, 1992.

\bibitem{LiChen2013} Y. Li, W. Chen, Y. Wang, Z.L. Zhang.  Influence diffusion dynamics and influence maximization in social networks with friend and foe relationships. In {\em Proceedings of the Sixth ACM International Conference on Web Search and Data Mining},  657--666, 2013.

\bibitem{Li2015} Y. Li, W. Chen, Y. Wang, Z.L. Zhang . Voter Model on Signed Social Networks. {\em Internet Mathematics}, 11, 93--133, 2015.

\bibitem{Moretti2013} P. Moretti, A. Baronchelli, M. Starnini, and R. Pastor-Satorras. Generalized Voter-Like Models on Heterogeneous Networks. In {\em Dynamics on and of complex networks}, Vol 2 (Springer New York), 285–300, 2013.


\bibitem{NazTag2012} A. Nazemian, F. Taghiyareh.  Influence maximization in independent cascade model with positive and negative word of mouth. In {\em Proceedings of the 6th International Symposium on Telecommunications (IST)},  pp. 854--860, 2012.

\bibitem{Shi2016} G. Shi, A. Prouti{\`{e}}re, M. Johansson , J. S. Baras and K.H. Johansson. The Evolution of Beliefs over Signed Social Networks. {\em Oper. Res.}, 64(3), 585--604, 2016.

\bibitem{Sirbu2017}
A. S{\^{\i}}rbu, V. Loreto, V.D. P. Servedio, and F.Tria. Opinion Dynamics: Models, Extensions and External Effects. In {\em Participatory Sensing, Opinions and Collective Awareness} (V. Loreto, M. Haklay, A. Hotho, V. D. P. Servedio, G. Stumme, J. Theunis, and F. Tria eds.), {\em Understanding Complex Systems} series, Springer, 363--40, 2017.
  
\bibitem{Stitch2014} L. Stich, G. Golla, A. Nanopoulos.  Modelling the spread of negative word-of-mouth in online social networks. {\em J. Decis. Syst.} 23(2), 203--221, 2014.

\bibitem{Xue2020} Xue Lin, Qiang Jiao and Long Wang. Competitive diffusion in signed social networks: A game-theoretic
perspective. {\em Automatica}, 12, 108656, 2020.

\bibitem{YaWaTru2019} S. Yang, S. Wang, V.A. Truong.  Online Learning and Optimization Under a New Linear-Threshold Model with Negative Influence. {\em arXiv}, arXiv:1911.03276. Available online: \texttt{https://arxiv.org/pdf/1911.03276.pdf}, 2019 (accessed on 27 January 2020).
\end{thebibliography}




\end{document}